\documentclass[12pt]{article}

\usepackage{amssymb}
\usepackage{graphicx}
\usepackage{amsmath}

\usepackage{times}

\newenvironment{sciabstract}{%
\begin{quote} \bf}
{\end{quote}}

\title{Artificial Rheotaxis}

\author
{
J\'er\'emie Palacci$^{1,6,\ast}$, Stefano Sacanna$^{2}$, Anais Abramian$^{3}$, J\'er\'emie Barral$^{4}$, \\ Kasey Hanson$^{5}$, Alexander Y. Grosberg$^{1}$, David J. Pine$^{1}$, Paul M. Chaikin$^{1}$\\
\\
\normalsize{$^{1}$ Department of Physics, New York University, USA,}\\
\normalsize{$^{2}$ Department of Chemistry, New York University, USA,}\\ 
\normalsize{$^{3}$ D\'epartement de Physique, Ecole Normale Sup\'erieure de Lyon, France,}\\
\normalsize{$^{4}$ Center for Neural Science, New York University, USA,}\\ 
\normalsize{$^{5}$ School of Materials Science and Engineering of Engineering, Geogia Tech, USA,}\\
\normalsize{$^{6}$ Department of Physics, UC San Diego, USA}
\\
\normalsize{$^\ast$To whom correspondence should be addressed; E-mail:  palacci@ucsd.edu.} 
}

\date{}

\begin{document} 


\baselineskip24pt


\maketitle


\begin{sciabstract}
{Motility is a basic feature of living microorganisms, and how it works is often determined by environmental cues. Recent efforts have focused on developing artificial systems that can mimic microorganisms, and in particular their self-propulsion. Here, we report on the design and characterization of  synthetic self-propelled particles that migrate upstream, known as positive rheotaxis. This phenomenon results from a purely physical mechanism involving the interplay between the polarity of the particles and their alignment by a viscous torque. We show quantitative agreement between experimental data and a simple model of an overdamped Brownian pendulum.  The model notably predicts the existence of a stagnation point in a diverging flow. We take advantage of this property to demonstrate that our active particles can sense and predictably organize in an imposed flow.  Our colloidal system represents an important step towards the realization of biomimetic micro-systems with the ability to sense and respond to environmental changes.}
\end{sciabstract}


\section*{Introduction}

In recent years, there has been a significant effort to design and synthesize active colloidal particles that mimic microorganisms. This has led to the development of self-propelled particles that can harvest a source of energy from their surroundings and convert it into directed motion \cite{Wang:2013cc, Sengupta:2012ie,Wang:2012kd, Buttinoni:2012ho}. These systems share many properties with their biological  counterparts: they exhibit a  persistent random walk analogous to  the run-and-tumble motility of bacteria, spermatozoa or algae \cite{Cates:2013ia} and some synthetic realizations can even  interact through chemical gradients and form "colonies", analogous to chemical-sensing  organisms \cite{Theurkauff:2012jo, Palacci:2013eu, Keller:1971vf,Masoud:2014es}.

A key feature of motility is its interaction and response to its environment. The ability to sense its surroundings is crucial for  living systems and the migration up or down a gradient is  called {\it taxis}. Phototaxis of invertebrate larvae contributes to the vertical migration of marine plankton, which is thought to represent the biggest biomass transport on Earth\cite{Jekely:2008kn}. 
 {\it Chemotaxis}, the migration in a chemical gradient, is observed  for (sea urchin) spermatozoa to guide themselves towards the egg \cite{Friedrich:2007vr} and  {\it E. Coli} bacteria to locate nutrient rich environments. It is furthermore argued to be one of the key component in the formation of patterns and colonies of bacteria \cite{Budrene:1995ew,Budrene:1991gq,Cates:2010um}.

The presence of  boundaries or obstacles can dramatically impact the motion of motile microorganisms. Motile bacteria like {\it E. Coli} move in circle near surfaces \cite{ISI:000258184500062,DiLeonardo:2011cpa} and aggregate on substrates to form biofilms,  at the root of many persistent bacterial infections \cite{Costerton:1999wa}.  They reverse directions when spatially constricted, and migrate preferentially through an array of V-shaped funnels \cite{PeterGalajda12012007}.  Additional effects emerge in a shear flow, e.g. a flow passed a no-slip solid surface, and  numerous biological organisms exhibit {\it rheotaxis} or migration under shear. Rheotaxis of fish arises from a complex biological sensing \cite{ARNOLD:1974bf,Montgomery:1997bo}, but sperm \cite{Bretherton:1961hf,Pedley:1992wa} and bacteria rheotaxis originate from physical mechanisms, such as a subtle interplay between velocity gradients and the helical shape of the flagella \cite{Anonymous:2012cu}.  {\it E. Coli} bacteria  as well as sperm cells exhibit direct and continuous upstream motility (positive rheotaxis) under shear \cite{Kaya:2012ch, Kantsler:2014ee}. Positive rheotaxis is also reported for {\it P. aeruginosa} parasites, as a coupling of the flow alignement with the twitching motility of the organism. This ability to go against the flow is argued to be beneficial for the ability to colonize environments, like the bladder, that are inaccessible to other human pathogens \cite{Shen:2012gg}. In these cases, the upstream migration is hypothesized to result from the coupled effect of the alignment of the body with the shear and the polar propulsion of the organism \cite{Shen:2012gg,Kaya:2012ch,Hill:2007et,Meng:2005ez, Kantsler:2014ee} resulting in migration towards the organs and areas that flows are flushing.

Here we present a biomimetic colloidal system consisting of particles that exhibit a continuous positive rheotaxis under flow, near a solid substrate. Their upstream migration originates from a flow-induced alignment of the particles' polar heads. This restoring hydrodynamic torque  competes with the thermal noise to align the particle and the system is quantitatively described by an overdamped Brownian pendulum in an effective potential induced by the flow.  We show that this mechanism allows for the development of adaptive particles that react to an environmental change. 

Synthetic rheotactic particles constitute a step towards the engineering of advanced microsystems with sensing, analogous to microorganisms.

\section*{Results} 
Our new particles  are engineered starting  from the colloidal surfers presented in \cite{Palacci:2013eu}: composite particles featuring a hematite cube embedded in a polymeric sphere.

We showed in a previous study that these particles become active under activation by a blue light in the presence of hydrogen peroxide fuel. They do not swim in bulk and only self-propel on a substrate in the considered experimental conditions (pH$\sim8.5$). In a nutshell, the blue light triggers the photocatalytic decomposition of the hydrogen peroxide fuel by the hematite, creating chemical (O$_2$ and H$_2$O$_2$) gradients. Phoresis and osmosis are  complementary interfacial phenomena \cite{anderson}. In a gradient, a free colloid  migrates phoretically; alternatively, placed in the same gradient, a fixed surface will develop an osmotic flow in the opposite direction. Consequently, under activation of the hematite by light, the surface of the substrate develop an osmotic  pumping flow, due to the concentration gradient produced by the decomposition of the fuel\cite{CordovaFigueroa:2008db,janus_propulsion_ajdari}. The hematite part of the particle is pulled toward the substrate by the osmotic flow and the particle propels, surfing over the osmotic flow \cite{Palacci:2013eu}.

Using a thermal treatment [see Materials and Methods], we make the hematite inclusions to protrude out of the particle polymer matrix, obtaining the anisotropic particles shown in [Fig. 1A].
In a typical experiment the particles are dispersed in a water-based solution containing hydrogen peroxide (varying concentration 1-10\% v/v) and $5~\rm{mM}$ tetramethylammonium hydroxide to fix the pH, pH$\sim8.5$. 
The suspensions are contained inside borosilicate square capillaries (Vitrotubes,  $500~\mu$m) that were previously cleaned by oxygen plasma. The capillaries are connected to a syringe pump (Harvard apparatus 33) using standard microfluidic tubing (Peek tubing, $125~\mu$m inner diameter) and sealed using UV curing glue (Nordland 86). A controlled and steady flux of solvent can be  imposed in the capillary thanks to this setup. The experiments are observed through an inverted microscope (Nikon TE300) and recorded using a monochrome camera (EO-1312M)  at a framerate in the range 20-30fps.

Under bright-field illumination, the particles are at equilibrium with the solvent and sit at a gravitational height $h_g\sim 100$nm from the  bottom wall [see Materials and Methods].  The particles exhibit thermal translational and rotational brownian motion, visible because of the optical contrast provided by the red hematite cube.   When illuminated through the microscope objective (oil immersion, 100x, N.A.=$1.4$) with blue light  (Nikon Intensilight, equipped with a bandpass filter $\lambda\sim430-490\rm{nm}$), the composite particles start to self-propel, with a distinct head and tail. The hematite cube leads the self-propelled particle, consistent with observations made using bare hematite particles as colloidal dockers for cargo transportation \cite{Palacci:2013tu} and previous experiments with spherical composite particles directed by a magnetic field \cite{Palacci:2013eu}. 

In the absence of an imposed flow, the particles exhibit isotropic self-propulsion with average velocity $V_0$ [Fig 1B-inset; movie S1], which  varies   with the concentration of hydrogen peroxide fuel and follows Michaelis-Menten kinetics [Fig. 1B] \cite{Michaelis:1913um, Howse:2007ed}. The velocity linearly increases with hydrogen peroxide concentration at low concentrations and saturates at $V_0\simeq 8\mu$m/s for [H$_2$O$_2$] $\gtrsim 5\%$ w/w.   

  The flux-driven flow is imposed in the $x$ direction by a syringe pump connected to the capillary tube. This creates a steady Poiseuille flow so that the particles, confined near the non-slipping bottom surface, experience a shear flow. In the following, we denote $V$ the velocity of the particle and $v$ is the velocity of the fluid at the particle's center.  The velocity of the flow at the altitude $y$ is $v(y)=\dot{\gamma}\ y$ where $\dot{\gamma}$ is the local shear rate [Fig. 1C]. 

In the case of a point particle advected by the flow, the two velocities are equal $V(y)=v(y)$. For the general case of a particle with finite radius $R$, the shear flow exerts a torque on the particle which exhibits translational as well as rotational motion, and the ratio of the two velocities  is a function of the dimensionless number $y/R$, where $y$ is the distance of the center particle to the wall, $V(y)/v(y)=f(y/R)$. The function $f$ is an increasing function of the ratio $y/R$ and $f(y/R)\rightarrow 1$ far from the wall, $y\gg R$ \cite{Goldman:2001ur}.

Without activation by light, the particles are passive tracers of the flow. They translate while rolling, on average, at the altitude $y_g=R+h_g$ [Fig. 1C].   The streamlines are straight lines along the $x$ direction  [Fig. 1D-E dashed line] and we characterize the flow measuring $V_x$. In the following, we use the notation  $^*$ for the  velocity of the particles measured in the absence of activation by the light, $V^*=V_x= f(y_g/R) \times v(y_g)$.    

The dynamics of the particles in an external flow  is drastically altered if the light is turned on and the self-propulsion activated [Fig 1D-E, solid lines]. The particles stop rolling, and make a turn so that the hematite protrusion faces the imposed flow [Fig 1D, solid line; movie S2, S3]. For moderate flow, $V^*\leq V_0$, the particles migrate upstream once activated [movie S2]. Increasing the flow velocity results in a better alignment of the particle displacement along the flow [movie S3]. The angular probability distribution of the displacement $P(\theta)$ is flat for small flows, $V^*=0.6~\mu$m/s,  and narrows down as the flow velocity increases, leading to an acute distribution for  $V^*=20~\mu$m/s [Fig. 2C]. 

We  do not observe any dependence of the phenomenon along the transverse position $z$ in the channel.
 For fast flows, $V^*\geq 35~\mu$m/s, the particles  detach from the wall due to the high shear. This regime is not discussed in the paper.
 
 In a set of independent experiments, we test the effect of the rotational motion of the particle on the translation velocity. The particle is advected by the flow and, in the absence of any light, rotates. We can freeze the rotation of the particle using a uniform magnetic field to set the direction of the magnetic moment of the hematite. We do not observe any modification of $V$, whether the field it is applied transversely or along the flow. This result is  consistent with  \cite{Goldman:2001ur} and shows that translation and rotation are decoupled at the altitude $y_g$.

In the following, we use bracket notation $\langle \rangle$ for ensemble and time averages, where a typical ensemble includes 6 to 16  independent  particles. 
We use the particle velocity $\langle V_x \rangle$ to quantify the experimental results. We measure $\langle V_x \rangle$ for various flows velocity  $V^*$  and for various self-propulsion velocities $V_0$, obtained by varying the concentration of hydrogen peroxide [Fig. 2A]. 

In the absence of flow, the propulsion is isotropic and the particles have equal probability to go against or along the flow, $\langle V_x \rangle =0\mu$m/s. Negative values for $\langle V_x \rangle$ indicate positive rheotactic behavior of particles  migrating upstream. For large flow velocities, $\langle V_x \rangle$ increases linearly with $V^*$.

\section*{Discussion}
 Our understanding of the rheotaxis in this system has to do with how the active polar particle aligns with the flow.  Under light activation, the hematite cube is pulled towards the surface due to the induced osmotic flow at the wall \cite{Palacci:2013eu,Palacci:2013tu}, the polymer body lags behind the center of the particle, and the viscous drag exerts a torque ultimately resulting in the alignment of the particle by the flow. The polarity of the self-propelled particle induces the upstream migration.  
The osmotic pumping additionally  pulls the particle closer to the wall, at the altitude $y_{on}=R+h_{on}$, where $h_{on}$ is typically a few Debye length $\lambda_D\sim 4$nm. The greater proximity of the non-slipping wall slows down the particles: $V(y_{on})= f(y_{on}/R) \times v(y_{on})$ with $f(y_{on}/R)< f(y_{g}/R)$. We express $V(y_{on})$, the advected velocity, in terms of the measurement $V^{*}$: 
\begin{equation} 
V(y_{on}) \sim \frac{f(y_{on}/R)}{f(y_g/R)} V^*
\label{V_v}
\end{equation}

We now present a quantitative description based on this simple model.   We use  the  bold notation for vectors and denote by $\eta$ the viscosity of the solution, equal  to the viscosity of water to a very good approximation. We assume, for the sake of simplicity, that, under light activation, the hematite cube is a {\it fixed} pivot, and we use the notation presented on [Fig. 2B] .

The flow exerts on the particle a Stokes drag ${\bf F}_\mathrm{S}=6\pi \eta R {\bf v}$ leading to a torque ${\bf M}_\mathrm{S}=6\pi \eta R^2 v \sin \theta \ {\bf e}_z=M_\mathrm{S}\   {\bf e}_z $, which tends to align "downstream" from the hematite pivot  [Fig. 2B] .
There is a competition between the restoring torque and the thermal rotational diffusion  of the particle around the hematite pivoting point. Provided the low Reynolds dynamics of the particles in the experiment, $Re\sim 10^{-5}$, the problem is formally equivalent to an overdamped Brownian pendulum in a "gravitational" field, the gravity being here replaced by an effective potential \cite{Barrat:2003wc}
\begin{equation}
\label{eq:Ueff}
U_{\rm{eff}}(\theta)=-6\pi \eta R^2 v \cos \theta 
\end{equation} 
 Using this analogy, the  equation followed by the angular probability distribution   P$(\theta,t)$ is given by the Smoluchovsky  equation \cite{Barrat:2003wc}: $ \partial_t P(\theta,t)=D_\theta \partial_\theta [\partial_\theta P(\theta,t) + \beta M_S P(\theta,t)]$ where $D_\theta$ is the rotational diffusion coefficient of the particle,   $\partial_\theta$ the angular component of the gradient, and $\beta=1/k_BT$. 
\\
The steady state solution is given by the pseudo-Boltzmann distribution $P(\theta)\propto \exp(-\beta U_{\rm{eff}})$. After normalization, and using the expression of the effective potential $U_{\rm{eff}}$ [Eq. \ref{eq:Ueff}], we obtain: 
\begin{equation}
\label{eq:P}
P(\theta)=\frac{1}{I_0 (K)} e^{  K cos \theta}
\end{equation}
where $I_0$ is the modified  Bessel function of the first kind of $0^{th}$ order \cite{Wolfram:1999vr} and $K=6\pi \eta \beta R^2 v=v/\tilde{v}$. In our present experimental conditions, we estimate from the former equation, $\tilde{v}\sim 0.2~\mu$m/s.

We compare the theoretical expectations for the {\it orientation} distribution function of particle [Eq. \ref{eq:P}] with the experiment. We measure experimentally the angular distribution of the {\it  displacement} as the angle made by the displacement of the particles, with respect to the direction $x$ of the flow, over a time interval of  $\Delta t=0.2$s, much shorter than the typical persistence time of self-propulsion $\sim 8s$, determined from the trajectories.  Given the polarity of the moving composite  particle, this measurement reflects the orientation of the particle, while being much more experimentally robust and accurate than the determination of the orientation vector of the particle. The distributions are obtained for  $100-600$ angles per particle for $\sim 10$ independent particles. 
 The results show good agreement between the pseudo-Boltzmann distribution and the experimental measurements [Fig. 2C]. We use the distributions to extract the dependance of the width $K$  as a function of the particle velocity $V^*$.  The datas agree, over two decades,  with the  linear scaling predicted by our model, $K= V^*/\tilde{v}_{\mathrm{exp}} $    with $1/\tilde{v}_{\mathrm{exp}}=1.5\pm0.5\ \mathrm{s/\mu m}$ [Fig 2C-inset]. One should note that the model assumes that the hematite cube provides a rigid fixed pivot while we observe in the experiment that it is sliding on the substrate. As a consequence, we expect to overestimate the strength of the restoring torque from the model, in line with the observed result.
 
Given the angular distribution $P(\theta)$, one can derive the expression for the average velocity $\langle V_x\rangle$. In our simple model, the instantaneous velocity of the particle in the reference frame of the lab is given  by  $V_x=V(y_{on}) - \cos\theta \ V_0$, where $V(y_{on})$ is the particle velocity at the altitude $y_{on}$ and $-\cos\theta \ V_0$ is  the contribution of the of the self-propulsion, due to the alignment of the particle by the flow. Using [Eq.\ref{V_v}], we average over both time and an ensemble of  independent particles and obtain: 
\begin{equation} \begin{split} \left< V_x \right> & = \left< \frac{f \left( y_{\mathrm{on}} /R \right)}{ f \left(y_g/R \right)} V^{\ast} - V_0 \cos \theta \right> =  \frac{ f\left(y_{\mathrm{on}}/R \right)}{f \left(y_g/R \right)} V^{\ast} - V_0 \left< \cos \theta \right> = \\
 & = \alpha V^{\ast} - \frac{I_1 \left( K \right)}{ I_0 \left( K \right)} V_0 \label{eq:final} \end{split}
\end{equation}
 where $I_1=\int P(\theta)\cos \theta \mathrm{d} \theta$ is the modified  Bessel function of the first kind of $1^{st}$ order \cite{Wolfram:1999vr} and $\alpha=\frac{f(y_{on}/R)}{f(y_g/R)}$ is smaller than 1.

 We compare this simple model with the experimental measurements gathered on Fig. 2A using the value for $\tilde{v}$ extracted from  the experiment [Fig. 2D] and the velocity of self-propulsion, $V_0$, measured in the absence of  flow [Fig. 1B].  We obtain an excellent quantitative agreement between the model provided by [Eq. \ref{eq:final}] [dashed lines] and the experiment [data points] as visible on Fig. 2, using $\alpha\sim0.75$ for all the sets of self-propulsion and hydrogen peroxide concentration (1\%, 3\% and 10\% v/v). This value for $\alpha$ corresponds to a change of height from $h_s\sim100$nm  to $\sim10$nm according to Goldmann {\it et al} \cite{Goldman:2001ur}, and in good agreement with our  estimates for the change of altitude of the particles, under activation of the particles by the light and change from the equilibrium sedimentation height to a few Debye length. \\
A similar mechanism has been discussed for mammalian sperm cells  and compared to a dynamical model using resistive force theory. Sperm cells spontaneously swim towards surfaces, heads on. The tails therefore explore, on average, regions of higher flow velocity than the head, resulting in a net torque and shear-induced rectification \cite{Kantsler:2014ee}.\\
 
The observed phenomenon exhibits an interesting feature: the existence of a stable stagnation position for the particles in a diverging flow field  [Fig. 2A]. We harness this property to design a system of adaptive active particles which self-organize in a non-uniform shear flow. We obtain a diverging flow placing  the tip of a pulled micropipette a few microns above a glass substrate, immersed in a solution  rheotactic particles and hydrogen peroxide [Fig. 3-insets].  The flow velocity and the shear is decaying from the nozzle. In the absence of activation by light, the flow flushes the particles away from the pipette. Once activated by the light, the particles spontaneously gather along a circle around the tip, at a finite distance $\Delta R_0$ from the tip and determined by the stagnation velocity $\langle V_x\rangle =0$ for non zero $U^*$ [Fig. 3-insets and movie S4]. Turning off, the light, the particles are immediately flushed away from the tip with a velocity of $7~\mu$m/s in agreement with the model [Fig 3]. They reorganize and reform the circle at the same distance   $\Delta R_0$ once the light is reactivated, the flow being unchanged. This is a step toward the  achievement of an adaptive artificial system which organizes in  an external flow.
\\

 In this paper, we engineered synthetic micro particles which can sense their environment and spontaneously migrate against an imposed external flow near a surface, thus exhibiting a rheotactic behavior. This effect originates in the alignment of the body of a polar self-propelled particle by a drag anisotropy. It can result from a shape anisotropy or the attraction of one part of the body to a solid substrate (rheotaxis on a surface), as for mammalian sperm cells \cite{Kantsler:2014ee} or our particles.  Our system is formally  identical to an overdamped Brownian pendulum in an effective non-equilibrium potential, induced by the flow, and is quantitatively described by a Schmoluchovsky equation for the angular distribution. This system shows a step forward in the synthesis of artificial microsystems with advanced biomimetic functionalities, since analogous strategies are used by many microorganisms to colonize regions against the flow \cite{Shen:2012gg,Kaya:2012ch,Hill:2007et,Meng:2005ez}. 
 
 \section*{Materials and Methods}
\subsection*{Colloidal synthesis and Heat treatment}
The photo-catalytic hematite cubes were prepared following the method described by Sugimoto {\it et al.} \cite{Sugimoto:1993bf}. Briefly, a sodium hydroxide solution (21.6g NaOH in 100 mL of Milliopore  water) was dripped in an iron chloride solution (54g of $FeCl_3 \cdot 6H_2O$ in 100 mL of Milliopore  water) and the resulting gel was left undisturbed at 100 $^{\circ}$C in a closed Pyrex bottle for 8 days. The resulting hematite cubes were first washed by centrifugation and dispersed in deionized water and then encapsulated into polymerizable silicon oil droplets as follows.

First,  3-methacryloxypropyl trimethoxysilane (TPM)  was hydrolyzed to water-soluble silanols by vigorously mixing 5 mL TPM and 100 mL Millipore water until all the TPM was dissolved and a clear solution was formed (usually 5 to 7 hours). Then 0.75 mL of ammonia (NH$_3$, 37 \%wt) were added to a 30 mL aqueous dispersion of hematite cubes (final concentration ~10 \%wt) containing a total of 15 mL hydrolyzed TPM, while kept under vigorous magnetic stirring. The addition of ammonia causes a polycondensation reaction between metastable water-soluble silanols to produce insoluble silsesquioxanes that phase separate nucleating monodisperse droplets as further described in \cite{Rossi:2011gl}. Each hematite cube, initially suspended in the water phase, acts as a nucleation site for the formation of an oil droplet, thus, producing polymer droplets with a single inorganic inclusion. 
The reaction was sampled every 15 minutes and the particle growth was monitored with optical microscopy. To increase the size of the TPM droplets, the reaction was fed with more hydrolyzed TPM until the particles reached the desired size. The particles were then polymerized by adding 0.5 mg of 2,2'-azobisisobutyronitrile (AIBN) to the dispersion preheated at 80$^{\circ}$C and kept at this temperature for 3 hours. 
The polymerized particles were then dried in air at 80$^{\circ}$C to a solid powder and finally further heated at 700$^{\circ}$C for 5 hours. During this heat treatment the organic material contained in the TPM matrix burns causing a severe particles shrinkage and the consequent  protrusion of the hematite cubes.
  \subsection*{Sedimentation height}
The sedimentation height, h$_g=\frac{kT}{mg}$, is defined as the height for which the gravitational energy of the particle balances the thermal energy in the system. For a typical particle,  the typical radius of the TPM (density 2.0) sphere is $R\sim1\mu m$ and the hematite cube (density 5.25) of size $600$nm
 gives a sedimentation height for the composite particle, $h_g\sim100$nm.


\clearpage
\begin{figure}[htbp]
\begin{center}
  \includegraphics[width=.9\textwidth]{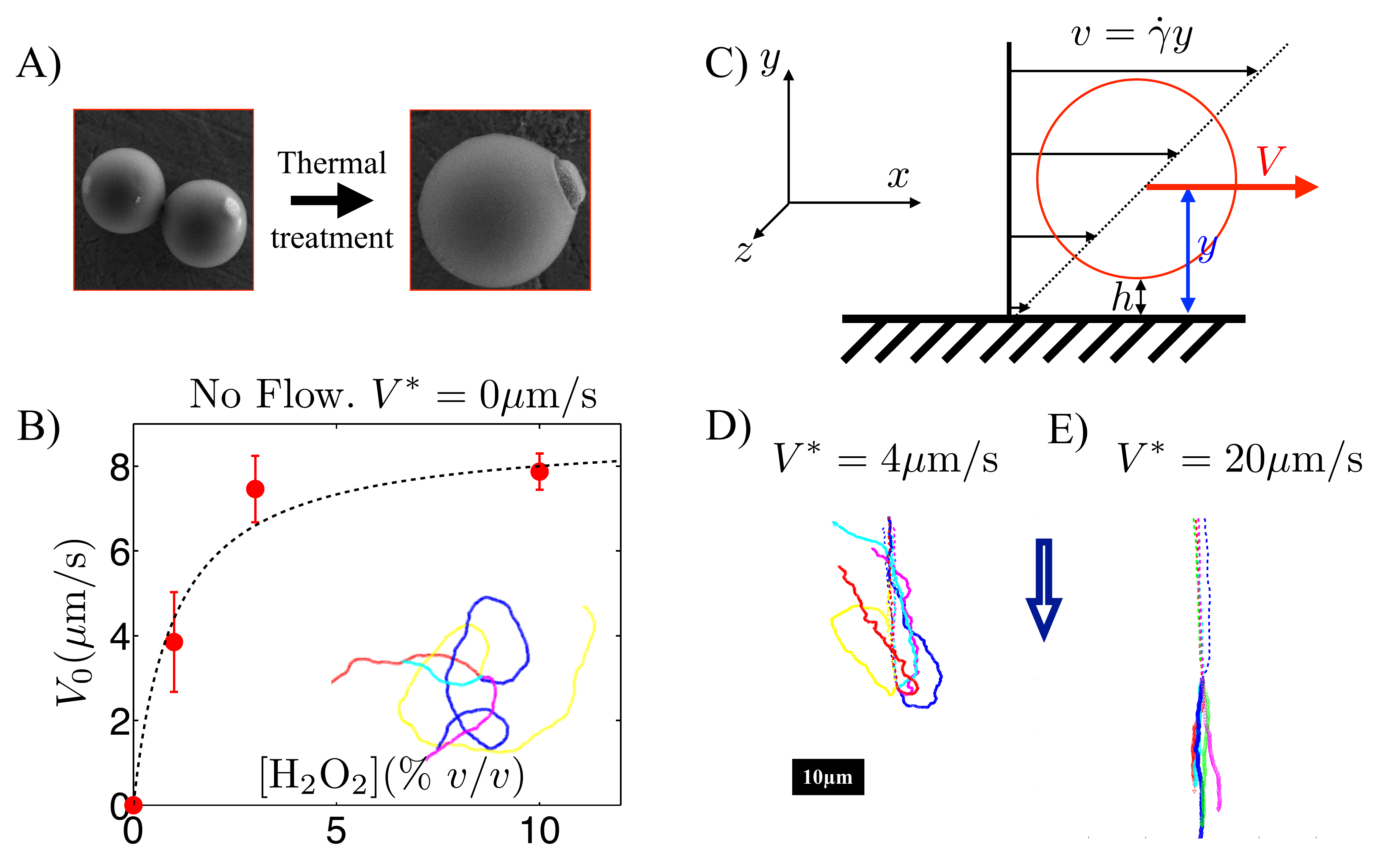}
\end{center}
\caption{(\textbf{A}) Spherical 3-methacryloxypropyl trimethoxysilane polymer (TPM) containing a hematite cube (left) undergo a thermal treatment, resulting in an anisotropic particle with the hematite cube protruding out (right).  (\textbf{B})  Dependence of the propulsion velocity $V_0$ with the concentration of hydrogen peroxide fuel  [H$_2$O$_2$]. The data is empirically fit (black dashed line) by a Michaelis-Menten Kinetics $V_0=V_{max} [H_2O]/(B+[H_2O_2])$.  (\textbf{B-inset}) Trajectories of different particles. We observe an isotropic propulsion  in the absence of any flow, and once activated by light. Note that the particles do not swim in bulk and only self-propel near a substrate. (\textbf{C}) Experimental setup. A syringe pump is connected to the capillary and induces a flow along the $x$ direction. The particle (red sphere)  reside at an altitude $y$ near the bottom surface. It experiences a shear flow $v=\dot{\gamma} y$, where $\dot \gamma$ is the local shear rate, near the non-slip boundary condition. This results in a translational velocity of the particle  $V(y/R)$ which depends on the ratio of the distance of the center of the particle to the wall with the radius $R$ of the particle. (\textbf{D}) Trajectories of the  particles for  various flows and $V_0\sim8~\mu$m/s.  The direction of the flow is indicated by the blue arrow.  In the absence of any light, the particle is used as a flow tracer advected at velocity $V^*$ (dashed line). Under light activation, the particle makes a U-turn and the hematite protrusion faces the flow  (solid line). The alignment  together with the polarity of the self-propulsion results in the upstream migration of the particles.}
\end{figure}

\clearpage
\begin{figure}[htbp]
\begin{center}
  \includegraphics[width=.5\textwidth]{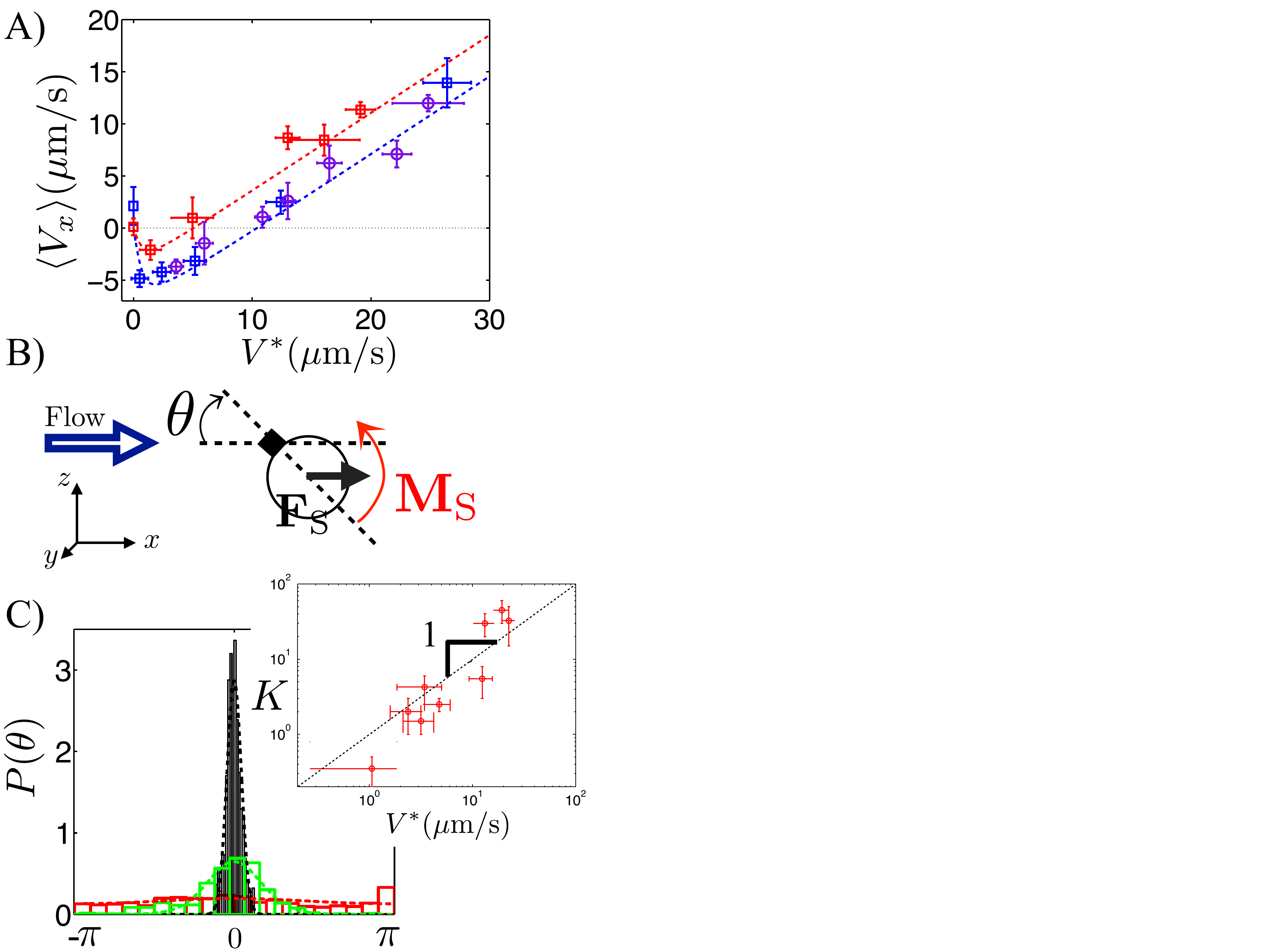}
\end{center}
\caption{(\textbf{A}) Projection of the average velocity $\langle V_x \rangle$ along the direction of the flow as a function of  $V^*$.  The measurements are performed for various hydrogen peroxide concentration  [H$_2$O$_2=1\%, 3\%, 10\%$ v/v] (respectively red, blue and purple symbols). The velocity $V_0$ depends on the fuel concentration. Fit of the experimental data with the model [see main text] for   $V_0=4~\mu$m/s (red dashed line) and $V_0=8~\mu$m/s (blue dashed line)  measured in the absence of any flow.  (\textbf{B}) Model for the artificial rheotaxis and notation. We assume that the hematite component acts as a fixed pivot, linked to the substrate. We denote by $\theta$ the angle made by the tail-head direction of the particle with the flow (see sketch). The flow exerts a viscous drag $F_s$ on the body of the particle, leading to  a torque $M_{F_s}$,  aligning the particle with the flow. The polarity of the particle induces an upstream migration.  (\textbf{C}) Normalized angular distributions P$(\theta)$ for increasing $V^*$  in the range $0.6$ to $20~\mu$m/s. Expected Boltzmann distribution  P$(\theta)\propto \exp[K \cos(\theta)]$ (dashed line) superimposed on the experimental histograms, with $K$ as the fitting parameter. (\textbf{C-inset}) Measurements of $K$ as a function $V^*$  [log-log scale]. The datas agree with  the linear scaling $K=V^*/\tilde{v}_{\mathrm{exp}}$ (black dashed line) predicted by our model. We obtain   $1//\tilde{v}_{\mathrm{exp}}=1.5\pm0.5$s/$~\mu$m from the fit to the experimental data. }
\end{figure}

\clearpage
\begin{figure}[htbp]
\begin{center}
  \includegraphics[width=.8\textwidth]{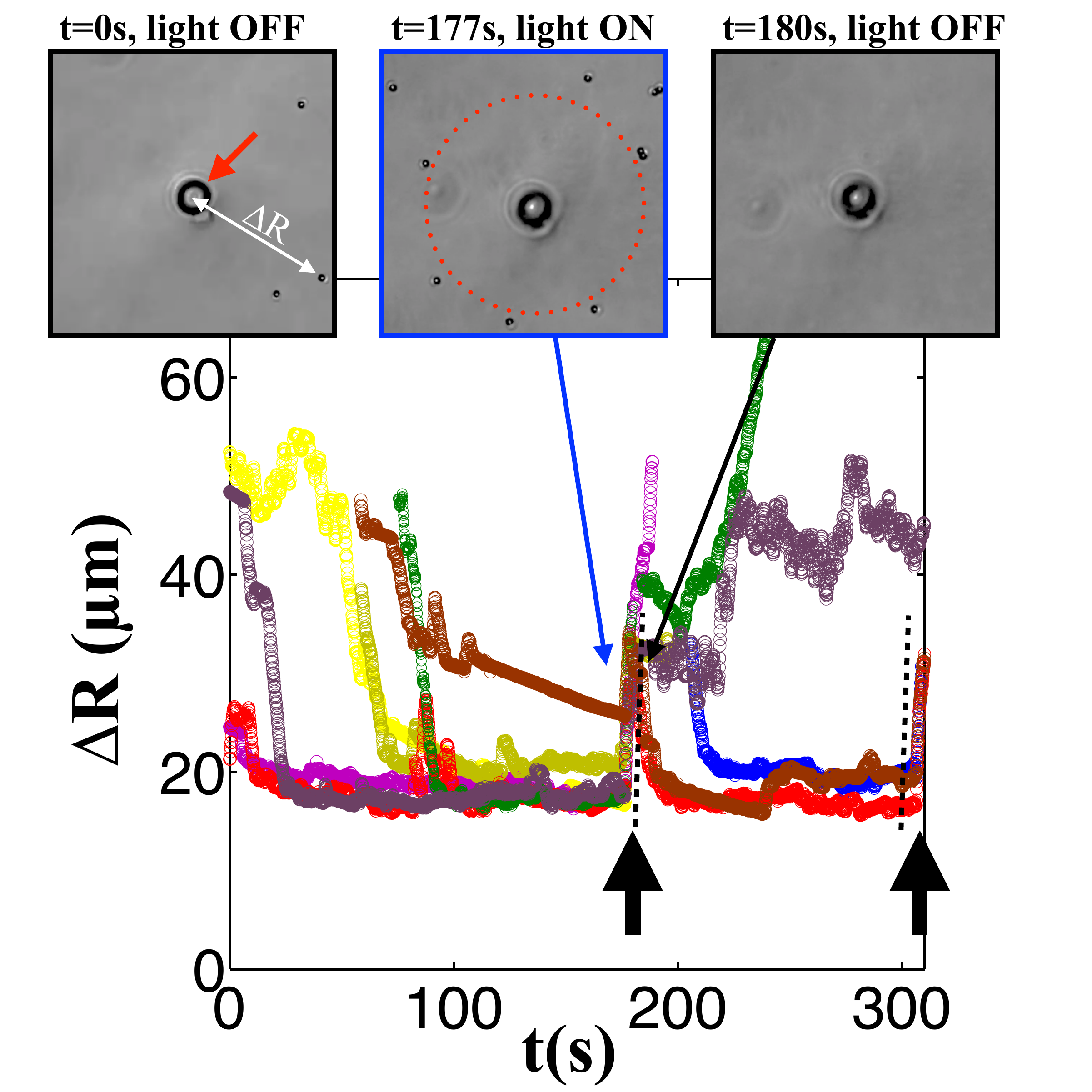}
\end{center}
\caption{{\bf Self-organization in a non-uniform flow field}. A pulled micropipette is approached from a glass substrate where reside artificial swimmers, dispersed in a fuel solution. The same fuel solution is ejected out of the nozzle of the micropipette (red arrow) inducing a diverging flow decaying from the tip. Initially, the particles are randomly distributed in the sample. Under light activation, the particles migrate towards the tip, and stop at a finite distance. They form a circle around the tip. Turning off the light, the particles are flushed away by the flow.  The behavior is quantified measuring the distance $\Delta R(t)$, as a distance to the nozzle (one color for each particle). The particles move towards the tip and stay at a finite and constant distance [Fig. 3-inset, t=177s]. Turning off the light (black arrows), they are flushed away by the flow [Fig. 3-inset, t=180s]. The velocity with which the particles are ejected provides a measure of the flow velocity at the position of mechanical equilibrium. We measure $U^*=7\pm1\mu m/s$ (black dashed line), in agreement with our expectations for the considered active particles. Turning on the light, the particles migrates against the flow and reform a circle at the same distance from the tip. }
\end{figure} 

\clearpage

\section*{Supplementary Information}
{\bf Movie S1: Dynamics of the self-propelled particles in the absence of any external flow and shear.} Under activation by a blue light, the particle self-propel along the bottom surface of a capillary with a distinct head and tail, the hematite cube heading. The particle is polar and the self-propulsion in on average isotropic. Control passive colloids, visible on the movie, exhibit Brownian motion and no drift showing the absence of flow.  Real time movie.
\\
{\bf Movie S2: Dynamics of the self-propelled particles under shear, at slow flow} ($V^*=4 \mu$m/s and the self-propulsion velocity of the particles $V_0=8\mu$m/s). 
Initially, the blue light is off and the particle (visible by the presence of the black hematite cube) is advected by the shear flow, tumbling near the bottom wall of the capillary. Under activation by the blue light, the particle makes a turn and start migrating upstream, exhibiting a rheotactic behavior. The black cube is facing the flow. Control passive colloids, visible on the movie,  are continuously advected by the flow. Real time movie.
\\
{\bf Movie S3: Dynamics of the self-propelled particles under shear, at fast flow}
($V^*=20\mu$m/s and the self-propulsion velocity of the particles $V_0=8\mu$m/s).
Initially, the blue light is off and the particle (visible by the presence of the black hematite cube) is advected by the shear flow, tumbling. Under activation by the blue light, the particle makes a turn, the black cube facing the flow. The particle is advected downstream but with a lower velocity than the flow itself. Control passive colloids are continuously advected by the flow. Real time movie.
\\
{\bf Movie S4: Organization of the particles in a diverging flow.}
A diverging flow is created from the nozzle of a pulled micropipette, placed a few microns above the substrate. Under activation by the light, the self-propelled particles migrate and form a circle around the nozzle at a finite distance from the tip, given by the stagnation point. Turning off the light, the particles are immediately flushed away by the flow. Reactivated by the blue light, the particle reposition at the same distance from the tip. The drift of the tip of the pipette is due a mechanical drift of the micromanipulator. Real time x10.

\end{document}